\documentclass[prc,aps,a4paper,groupedaddress,superscriptaddress,nofootinbib,showpacs
,preprintnumbers,twocolumn]{revtex4}
\usepackage{graphicx}
\usepackage{amsfonts}
\usepackage{amssymb}
\usepackage{amsmath}
\usepackage{natbib}
\usepackage{dcolumn}
\usepackage{bm} 
\newcommand{\bwt}{\begin{widetext}}
\newcommand{\ewt}{\end{widetext}}
\newcommand{\beq}{\begin{equation}}
\newcommand{\eeq}{\end{equation}}
\newcommand{\bea}{\begin{eqnarray}}
\newcommand{\eea}{\end{eqnarray}}
\begin{document}
\title{Effect of the $\delta$-meson on the instabilities of nuclear matter under strong magnetic fields}

\author{A. Rabhi}
\email{rabhi@teor.fis.uc.pt}
\affiliation{Centro de F\' {\i}sica Computacional, Department of Physics, University of Coimbra, 3004-516 Coimbra, Portugal} 
\affiliation{Laboratoire de Physique de la Mati\`ere Condens\'ee,
Facult\'e des Sciences de Tunis, Campus Universitaire, Le Belv\'ed\`ere-1060, Tunisia}
\author{C.~Provid\^encia}
\email{cp@teor.fis.uc.pt}
\affiliation{Centro de F\' {\i}sica Computacional, Department of Physics, University of Coimbra, 3004-516 Coimbra, Portugal} 
\author{J.~Da~Provid\^encia}
\email{providencia@teor.fis.uc.pt}
\affiliation{Centro de F\' {\i}sica Computacional, Department of Physics, University of Coimbra, 3004-516 Coimbra, Portugal} 

\date{\today}  
\begin{abstract}

We study the influence of the isovector-scalar meson on the spinodal instabilities 
and the distillation effect in asymmetric non-homogenous nuclear matter under strong magnetic fields, of the order of 
$ 10^{18}-10^{19}$ G. Relativistic nuclear models both with constant couplings (NLW) and
with density dependent parameters (DDRH) are considered.  A strong magnetic
field can have large effects on the instability regions giving rise to
bands of instability and wider unstable regions. 
It is shown that for neutron rich matter the inclusion 
of the $\delta$ meson increases the size of the instability region for NLW models 
and decreases it for the DDRH models. The effect of the $\delta$ meson on the
transition density to homogeneous $\beta$-equilibrium matter is discussed.
The DDRH$\delta$ model predicts the smallest transition pressures, about half the values obtained for NL$\delta$. 
\end{abstract}
\pacs{21.65.-f 26.60.Kp 26.60.-c 97.60.Jd} 
\maketitle

Neutron stars with very strong magnetic fields of the order of
$10^{14}-10^{15}$ G are known as magnetars \cite{duncan92,usov,pacz} and they are believed to be the sources of the intense gamma and X rays detected in 1979
(for a review refer to \cite{harding}). To date, 16 magnetars have been identified as short $\gamma$-ray repeaters or anomalous X-ray pulsars, although some are still unconfirmed candidates~\cite{sgr}. However, according to Ref.~\cite{kouve}, a fraction as high as 10\% of the neutron star population could be magnetars.
These neutron stars are hot, young stars, $\sim$ 1 kyear old. 
 In~\cite{duncan}, Thompson and Duncan have considered turbulent dynamo
amplifications in young neutron stars as a mechanism for generating the strong
magnetic fields.

In magnetars the crust is stressed by very strong forces which deform the
crust and may crack it. Once the surface cracks, the violent motions blast
particles along the magnetic fields, triggering
gamma rays and x-rays. In particular, scientists believe that the  giant burst
of energy experienced by the magnetar SGR 1806-20 in 2004 was triggered by a "starquake" in the neutron star's crust that caused a
catastrophic disruption in the magnetar's magnetic field. SGR 1806-20 is the
most magnetic object observed and has 
on the surface a magnetic field of intensity over 10$^{15}$ G \cite{sgr}.

In~\cite{pethick94} the authors have shown how to obtain both the moment of
inertia of a neutron star as well as the fractional moment of inertia of its
crust only  in terms  of the mass and the radius of the star. For the
fractional moment of inertia of the crust has an additional
dependence of the equation of state (EOS) which enters through the values of the pressure and the density at the crust core transition \cite{pethick94,link99}. The transition density enters just as a correction to the fractional moment of inertia, while the transition
pressure is the main EOS parameter which defines that quantity.  Using this result,
Link  \textit{et al.} \cite{link99} have obtained a  lower limit for the neutron star radius with a given mass, from the glitches occurring in the Vela pulsar and in other six
pulsars. This constraint will put severe restrictions on the acceptable  equations
of state of stellar matter if the radius and mass of a neutron star is measured.

Spin-up glitches have also been observed in all known persistent AXP
\cite{kaspi,dib}. However, different from the conventional low-field radio
pulsar, the glitches in AXPs are accompanied by a significant recovery of the
spin down rate of the pulsar. It would be interesting if information of the
crust and star properties could also be obtained from these glitches. 

It was recently shown in Ref.~\cite{aziz09} that a strong magnetic field, 
of the order of $10^{18}$- $10^{19}$G, has large effects on the instability 
regions of nuclear matter. Relativistic nuclear models both with constant couplings and
with density dependent parameters were considered. It was shown that a strong magnetic
field can have large effects on the thermodynamic spinodal instabilities zones giving 
rise to bands of instability and wider unstable regions. As a consequence, it was predicted
larger transition densities at the inner edge of the crust of compact stars
with a strong magnetic field. The direction of instability gives rise to a very strong 
distillation effect if protons occupy only partially a Landau level. However, 
for almost full Landau levels an anti-distillation effect may occur.

In this paper, which completes Ref.~\cite{aziz09},
we  study the influence of the isovector scalar meson 
on the low density instabilities of asymmetric nuclear matter under strong magnetic fields.
We  estimate the density and pressure at the transition from   the non-homogeneous  
to the  homogeneous phase in stellar matter  from the crossing of the EOS in
$\beta$-equilibrium matter with the spinodal  
and discuss the effect of the magnetic field on the direction of instability. 
We consider two kinds of relativistic mean-field  approaches:
 non-linear Walecka models (NLW) models with constant coupling parameters and
density dependent relativistic hadronic (DDRH) models with density-dependent coupling
parameters. The last models seem to give more realistic results at subsaturation densities~\cite{camille08}.
The inclusion of the $\delta$-meson brings to the isovector channel the same
symmetry existing already in the isoscalar channel with the meson pair $(\sigma,\omega)$
responsible for saturation in RMF models~\cite{kubis97,liu}. 
The presence of the $\delta$-meson 
softens the symmetry energy at subsaturation densities and hardens it above saturation density, giving rise to stable compact stars  with larger masses
\cite{mp04}. In \cite{camille08} it was shown that the instability region in the isovector direction becomes smaller if the $\delta$ meson is included.
Therefore, it is expected that in the presence of a strong magnetic field these differences become larger.

In the present paper we consider the NLW models  NL$\rho$ and NL$\delta$~\cite{liu} and  the DDRH models TW~\cite{tw} 
and DDRH$\delta$~\cite{gaitanos}. Only NL$\delta$ and DDRH$\delta$ include the $\delta$-meson.

For the description of the neutron star matter, we employ the standard mean-field theory (MFT) approach. 
A complete set of the equations and the description of the method can be found in Ref.~\cite{aziz08, aziz09}.
The Lagrangian density of TW~\cite{fuchs,tw} and DDRH$\delta$~\cite{gaitanos,abmp04} models reads:
\bea
{\cal L}&=&\bar{\Psi}_{b}\bigg(i\gamma_{\mu}\partial^{\mu}-q_{b}\gamma_{\mu}A^{\mu}- 
m_{b}+\Gamma_{\sigma}\sigma+\Gamma_{\delta}\vec{\tau}_{b}\cdot\vec{\delta} \cr
&-&\Gamma_{\omega}\gamma_{\mu}\omega^{\mu}-\frac{1}{2}\Gamma_{\rho}\tau_{3 b}\gamma_{\mu}\rho^{\mu}
-\frac{1}{2}\mu_{N}\kappa_{b}\sigma_{\mu \nu} F^{\mu \nu}\bigg )\Psi_{b} \cr
&+&\frac{1}{2}\partial_{\mu}\sigma \partial^{\mu}\sigma
-\frac{1}{2}m^{2}_{\sigma}\sigma^{2}+\frac{1}{2} \partial_{\mu}\vec{\delta} \,\partial^{\mu}\vec{\delta}
-\frac{1}{2}m^{2}_{\delta}\vec{\delta}^{2} \cr
&+&\frac{1}{2}m^{2}_{\omega}\omega_{\mu}\omega^{\mu}
-\frac{1}{4}\Omega^{\mu \nu} \Omega_{\mu \nu}  \cr
&-&\frac{1}{4} F^{\mu \nu}F_{\mu \nu}
+\frac{1}{2}m^{2}_{\rho}\rho_{\mu}\rho^{\mu}-\frac{1}{4}  P^{\mu \nu}P_{\mu \nu},
\label{lagran}
\eea
where $\Psi_{b}$ are the baryon ($b$=$n$, $p$) Dirac fields; $\sigma$, $\omega$,  $\rho$, and $\delta$ represent the scalar, vector, isovector-vector and isovector-scalar meson fields, which are exchanged for the description of nuclear interactions and $A^\mu=(0,0,Bx,0)$ refers to a constant external magnetic field along the z-axis. 
The nucleon mass and isospin projection for the protons and neutrons are denoted by $m_{b}$ and $\tau_{3 b}=\pm 1$, respectively. The mesonic and electromagnetic field strength tensors are given by their usual expressions: $\Omega_{\mu \nu}=\partial_{\mu}\omega_{\nu}-\partial_{\nu}\omega_{\mu}$, $P_{\mu 
\nu}=\partial_{\mu}\rho_{\nu}-\partial_{\nu}\rho_{\mu}$, and  $F_{\mu
\nu}=\partial_{\mu}A_{\nu}-\partial_{\nu}A_{\mu}$. The nucleon anomalous
magnetic moments (AMM) are introduced via the coupling of the baryons to the electromagnetic field tensor with $\sigma_{\mu \nu}=\frac{i}{2}\left[\gamma_{\mu},  \gamma_{\nu}\right] $ and strength $\kappa_{b}$ with $\kappa_{n}=g_n/2=-1.91315$ for the neutron and $\kappa_{p}=(g_p/2-1)=1.79285$ for the proton, respectively. The electromagnetic field is assumed to be externally generated (and thus has no associated field equation), and only frozen-field configurations will be considered. The density dependent strong interaction couplings are denoted by $\Gamma$, the electromagnetic couplings by $q$ and the nucleon, mesons masses by $m$. The density dependent coupling parameters are adjusted in order to reproduce some of the nuclear matter bulk properties using the following parametrization
\beq
\Gamma_{i}(\rho)=\Gamma_{i}(\rho_{sat})f_{i}(x),\quad i=\sigma, \omega, \rho,\delta
\label{gam1}
\eeq
where $x={\rho}/{\rho_{sat}}$, with
\beq
f_{i}(x)=a_{i}\frac{1+b_{i}\left(x+d_{i}\right)^{2}}{1+c_{i}\left(x+d_{i}\right)^{2}},
\quad i=\sigma, \omega,
\label{gam2}
\eeq
for TW,
\beq
f_{\rho}(x)=\exp\left[ -a_{\rho}(x-1)\right], 
\eeq
and,
\beq
f_{i}(x)=a_{i}\exp\left[ -b_{i}(x-1)\right]-c_{i}\left(x-d_{i}\right), \quad i=\rho, \delta
\label{gam4}
\eeq
for DDRH$\delta$, with the values of the parameters $m_i$, $\Gamma_{i}$, $a_{i}$, $b_{i}$,
$c_{i}$ and $d_{i}$, $i=\sigma, \omega, \rho, \delta$ given in Table~\ref{table1}. 

\begin{figure}[t]
\centering
\includegraphics[width=1.0\linewidth,angle=0]{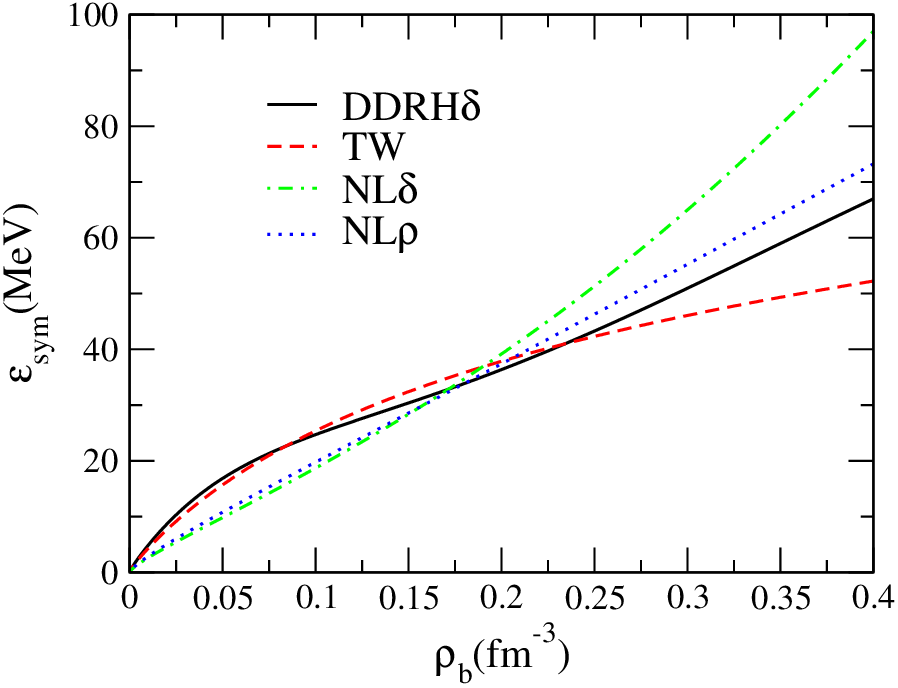}
\caption{Symmetry energy for all models used in the present work.}
\label{esym}
\end{figure}

In the sequel,  we define the magnetic 
field in units of the electron critical  field $B^c_e=4.414 \times 10^{13}$~G, so that  
$B=B^* \, B^c_e$.

For the DDRH$\delta$ model we use the parametrization  given in~\cite{gaitanos,abmp04} 
 except for the parameter $\Gamma_\rho(\rho_{sat})$ which we increase
so that the symmetry energy is 31 MeV at saturation density and not 25 MeV as in~\cite{gaitanos}, see Table \ref{table1}.
 The
inclusion of the $\delta$ meson reduces the symmetry energy at subsaturation
densities but makes it much stiffer at supra-saturation densities because the
$\delta$ meson field reaches saturation and the $\rho$-meson field
 increases always with density~\cite{kubis97,liu}. This is seen in Fig.~\ref{esym} where the 
symmetry energy  is given for the all models we are studying. 

\begin{table}[htb]
\begin{tabular}{ccccccc}
\hline
\hline
i &$m_{i}$& $\Gamma_{i}$ & $a_{i}$ & $b_{i}$ & $c_{i}$ & $d_{i}$  \\
& (MeV)&\\
\hline
$\sigma$ & 550 &10.72854 &1.365469 &0.226061 &0.409704&0.901995\\
$\omega$ &783 &13.29015 & 1.402488&0.172577 &0.344293&0.983955 \\
TW\\
$\rho$ 
 &763 & 7.32196 & 0.515 & & & \\
DDRH$\delta$\\
$\rho$
& 763 & 12.7530  & 0.095268  & 2.171 & 0.05336 & 17.8431 \\
$\delta$
&980 & 7.58963 & 0.01984 & 3.4732 & -0.0908 & -9.811 \\
\hline
\hline
\end{tabular}
\caption{Parameters for the TW and  DDRH$\delta$ models. These two models have the same parametrization for  the $\sigma$ and $\omega$ mesons.}
\label{table1}
\end{table}

For the NLW models, \textit{i.e.} NL$\rho$ and NL$\delta$, 
we add to the Langrangian density, Eq.~(\ref{lagran}), with $g_i=\Gamma_i$, the scalar meson self-interaction terms
$${\cal L}_{nl}=-\frac{1}{3}bm_n(g_{\sigma}\sigma)^3-\frac{1}{4}c(g_{\sigma}\sigma)^4,$$ where $b$ 
and $c$ are two dimensionless parameters. The coupling parameters are constant and given in Ref.~\cite{liu}. 

The stability conditions for asymmetric nuclear matter, keeping volume and temperature constant, are obtained from the free energy density $\cal F$, imposing that this function is a convex function of the densities $\rho_p$ and $\rho_n$, \textit{i.e.} the symmetric matrix with the elements~\cite{Bar03, marg03} 
\beq
{\cal F}_{ij}=\left( \frac{\partial^{2} {\cal F}}{\partial \rho_{i}\partial\rho_{j}}\right) _{T}
\eeq
is positive. At zero temperature, the free energy density coincides with the energy density. We define the thermodynamic spinodal at $T=0$ as the curve on the $(\rho_n, \rho_p)$ plane  for which the determinant of the ${\cal F}_{ij}$ is zero. 
The eigenvalues of the stability matrix are given by
\beq
\lambda_{\pm}=\frac{1}{2}\left(\hbox{Tr}({\cal F})\pm\sqrt{\hbox{Tr}({\cal F})^2-4 \hbox{Det}({\cal F})}\right) 
\eeq
and the eigenvectors $\delta {\bf \rho}_{\pm}$ by
\beq
\frac{\delta \rho^{\pm}_{i}}{\delta \rho^{\pm}_{j}}=\frac{\lambda_{\pm}-{\cal F}_{jj}}{{\cal F}_{ji}}, \: i, j = p, n.
\eeq

\begin{figure*}[htb]
\centering
\includegraphics[width=0.75\linewidth,angle=0]{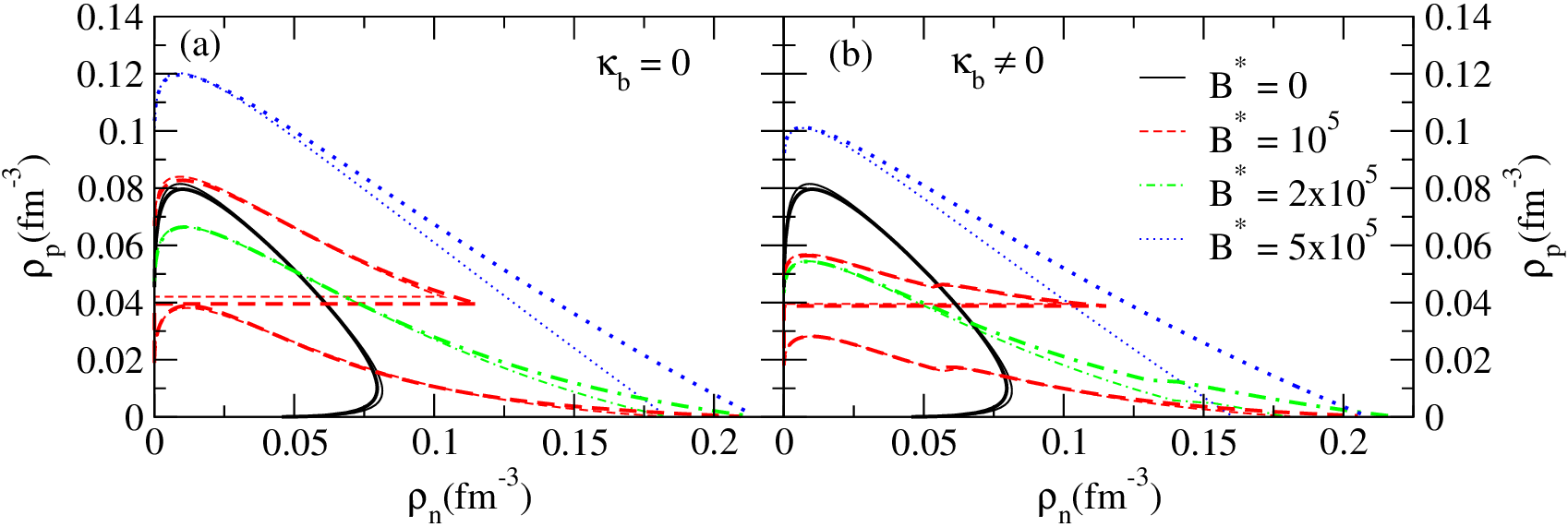}
\caption{(Color online) Spinodal section in terms of $\rho_p$ versus $\rho_n$ for NL$\rho$ (thin lines) and NL$\delta$ (thick lines) at $T = 0 \hbox{MeV}$ and for several values of magnetic fields (a) without and (b) with AMM.}
\label{spznlrdpn}
\end{figure*}

\begin{figure*}[htb]
\vspace{1.cm}
\centering
\includegraphics[width=.75\linewidth,angle=0]{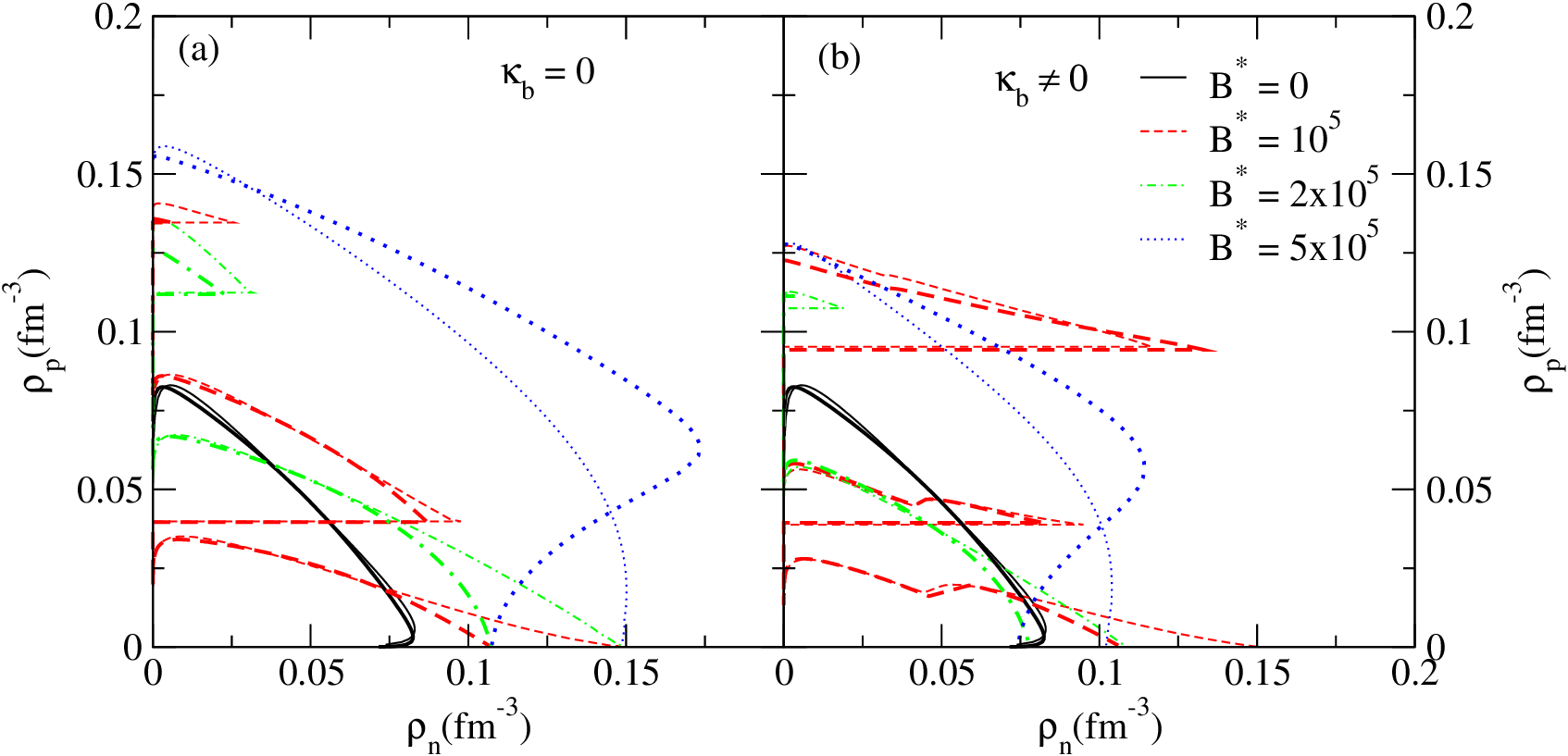}
\caption{(Color online) Spinodal section in terms of $\rho_p$ versus $\rho_n$ for TW (thin lines) and DDRH$\delta$ (thick lines) models at $T = 0 \hbox{MeV}$ and for several values of magnetic fields (a) without and (b) with AMM.}
\label{spzddrhdpn}
\end{figure*}

In Fig.~\ref{spznlrdpn} we show the spinodal sections 
for several magnetic field intensities and for two NLW models: NL$\rho$ without 
$\delta$-meson (thin lines) and NL$\delta$ including $\delta$ (thick lines). 
As discussed in~\cite{aziz09}, due to the Landau quantization,  the magnetic
field has a strong 
effect not only on the size, but also on the shape of the spinodal zone. Unlike the 
$B=0$ case, the instability zone is no longer symmetric with respect to the $\rho_n=\rho_p$ line. 
For the proton-rich matter, the inclusion of the $\delta$-meson has only a
small effect on the spinodal, namely a small reduction similar to the effect
already described for the $B=0$ \cite{pasta2}. For neutron-rich
matter in the presence of
a strong magnetic field, the effect of $\delta$-meson is stronger:
the NL$\delta$ model has a spinodal region larger than NL$\rho$.
This effect is similar whether the AMM is included or not.

The DDRH models behave in a different way due to density dependence of the
coupling parameters. In Fig.~\ref{spzddrhdpn}, we present the results 
of the thermodynamic spinodal instability region for TW (thin lines) and DDRH$\delta$ (thick lines). 
For $B^*=10^5\;\hbox{and}\; 2\times10^5$ the spinodal has three and two bands, see Fig.~\ref{spzddrhdpn}, 
corresponding to the occupation of the first three and two Landau levels, respectively.
For proton-rich matter with $B^*=10^5\;\hbox{and}\; 2\times10^5$, the inclusion 
of the $\delta$-meson reduces the size of the spinodal zone corresponding to 
the last occupied Landau level (LL), \textit{i.e.}, third LL for $B^*=10^5$ and second LL for $B^*=2\times10^5$. 
The other instabilities zones are not affected. 
For the neutron-rich matter, the $\delta$-meson reduces the size of the spinodal zones for all the magnetic fields considered, contrary to what happens with NL$\delta$.

For $B^* = 5\times10^5$ the protons are totally spin polarized in all the models under study and for the complete instability region. The size of the spinodal zone is the largest one.
\begin{table}[t]
\caption{
Predicted density, proton fraction and pressure at the inner edge of the
crust of a compact star at zero temperature, as defined by the crossing
between the thermodynamic instability region of $np$matter and the
$\beta$-equilibrium EOS for homogeneous, neutrino-free stellar
matter in the $(\rho_p,\rho_n)$ plane. The AMM is not included.} 
\label{table4}
\begin{ruledtabular}
\begin{tabular}{ccccc}
$B^{*}$ & Model & $\rho^{\hbox{cross}}_b (\hbox{fm}^{-3}) $ & $Y_p$ & $P_{m}(\hbox{MeV}\hbox{fm}^{-3}) $ \\
\hline
           $0$            & NL$\rho$          & 0.067 & 0.013 & 0.255 \\
                             &NL$\delta$        & 0.063 & 0.010 & 0.175 \\
                             &TW                    & 0.085 & 0.037 & 0.523 \\
                             &DDRH$\delta$   & 0.086 & 0.036 & 0.269 \\
$\phantom{..}$\\  
$3\times 10^{4}$           & NL$\rho$          & 0.071 &  0.043 & 0.284 \\ 
                           & NL$\delta$        & 0.071 & 0.040  & 0.264 \\ 
                           &TW                 & 0.071 &  0.056 & 0.342 \\ 
                           &DDRH$\delta$       & 0.077 & 0.056  & 0.259 \\ 
$\phantom{..}$\\  
$5\times10^{4}$& NL$\rho$          & 0.084 & 0.068 & 0.467  \\  
                             &NL$\delta$        & 0.084 & 0.064 & 0.467 \\
                             &TW                    & 0.082 & 0.086 & 0.461 \\
                             &DDRH$\delta$   & 0.084 & 0.086 & 0.306 \\
$\phantom{..}$ \\ 
   $10^{5}$         & NL$\rho$          & 0.105 & 0.121 & 0.822  \\
                             &NL$\delta$        & 0.105 & 0.117 & 0.896 \\
                             &TW                    & 0.101 & 0.146 & 0.673 \\
                             &DDRH$\delta$   & 0.096 & 0.148 & 0.404 \\
$\phantom{..}$\\
$2\times10^{5}$ & NL$\rho$         & 0.128 & 0.207 & 1.203  \\
                             &NL$\delta$        & 0.130 & 0.203 & 1.447 \\
                             &TW                    & 0.128 & 0.236 & 1.016  \\
                             &DDRH$\delta$   & 0.118 & 0.237 & 0.644 \\
$\phantom{..}$\\
$5\times10^{5}$ & NL$\rho$         & 0.162 & 0.367 & 1.275 \\
                             & NL$\delta$       & 0.170 & 0.369 & 1.830 \\
                             &TW                    & 0.198 & 0.402 & 2.331 \\
                             &DDRH$\delta$   & 0.228 & 0.422 & 4.460 \\
\end{tabular}
\end{ruledtabular}
\end{table}

The existing differences between the different models for neutron rich 
matter will affect  the transition density from a non-homogenous phase 
to a homogeneous phase in stellar matter. In fact, the  density  at the 
crossing of the EoS  for $\beta$-equilibrium stellar matter with the 
thermodynamic spinodal gives a reasonable prediction of the transition density 
\cite{bao,pasta1} of the crust to an homogeneous phase in stellar matter. 
It was shown in \cite{link99} that  the transition pressure
and, at a second level, the transition  density define the  fraction of the star's moment of inertia contained in the solid crust. In the sequel we will estimate for the different models and magnetic fields  the transition density and transition pressure at the inner edge of a compact star in $\beta$-equilibrium. We will discuss only the cases for which the crossing is occurring at the first LL. For fields  $B^*<3\times 10^4$ at  densities above the crossing with the first LL, there may occur other crossings with other LL. For these cases the determination of the extension of the non-homogeneous phase should be carried within  a more precise method. 

\begin{figure}[t]
\vspace{0.95cm}
\centering
\includegraphics[width=1.0\linewidth,angle=0]{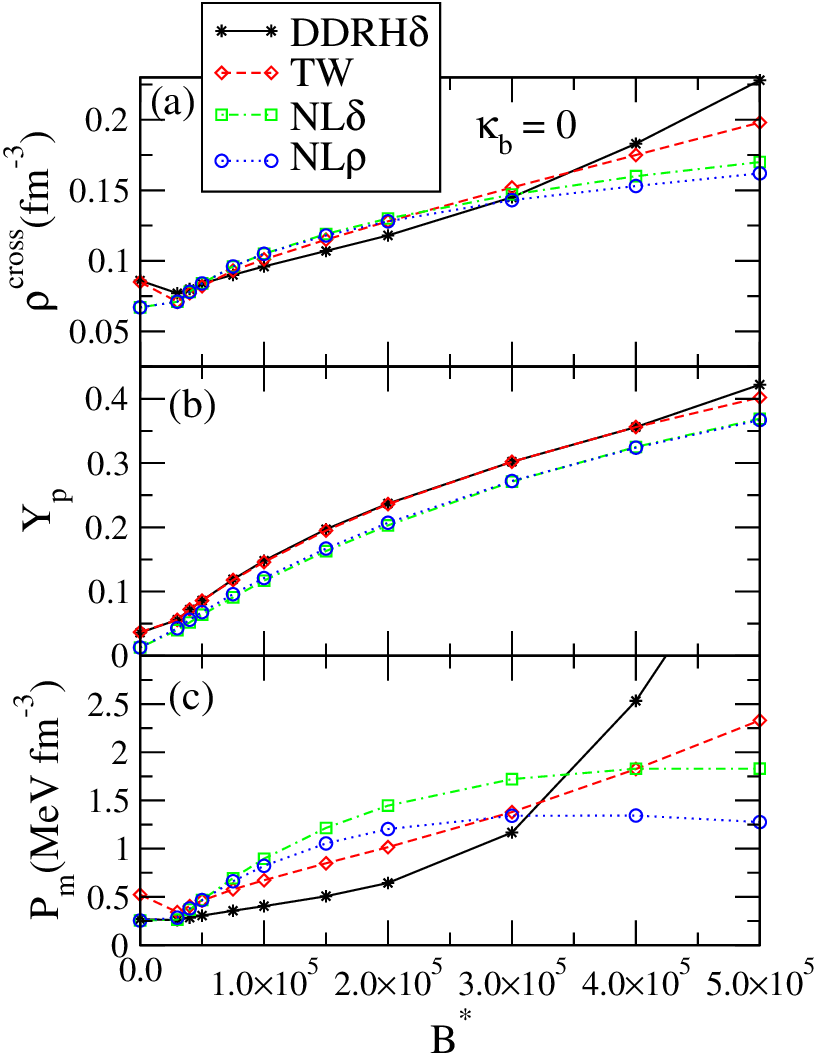}
\caption{(Color online) Physical quantities at the inner edge of a compact
star in the presence of strong magnetic fields: a) density, b) proton
fraction and  c) pressure.}
\label{denst}
\end{figure}

The values of the transition density  $\rho_b^{cross}$, and
respective proton fraction $Y_p$ and pressure $P_m$ are given for stellar matter under different magnetic
field intensities in Tables~\ref{table4} (without AMM)  and~\ref{table5} (with AMM). For convenience we have plotted the results without AMM  in Fig.~\ref{denst}.

For $B=0$, we can see that the inclusion of the $\delta$-meson reduces the transition density, and the corresponding proton fraction and  pressure for NLW models,  while  for DDRH models the transition density increases and the corresponding proton fraction and pressure decrease.
For DDRH models the crossing is occurring at three times larger proton fractions and twice the pressure.

For a finite magnetic field it was shown in~\cite{aziz09} for the models TW and TM1 (a parametrization similar to NL$\rho$) that the transition density increases with the magnetic field. We confirm this result when the $\delta$ meson is included and summarize the conclusions:\\
a) the
proton fraction at the crossing increases monotonically with the magnetic
field intensity, and for $B^*=5\times 10^5$ it reaches the values 0.36 and
0.4 respectively for NLW and DDRH models. The proton fraction does not depend on the inclusion
of the $\delta$ meson and NLW models predict systematically  a smaller proton
fraction; \\
b) the transition density is also increasing with the increase for $B^*\le5\times 10^4$. 
For smaller fields a decrease on the transition density may occur as it is seen with the 
DDRH models. The smaller the field, the larger the number of bands, defined by the 
filling of a new LL, will be crossed by the $\beta$-equilibrium EOS. As referred before, 
this situation requires a more carefully study.  For the smaller fields the transition 
densities for NL$\rho$ and NL$\delta$ are almost coincident and only above 
$B^*=2\times 10^5$ does the  NL$\delta$ predict a larger transition
density. Looking at the Fig.~\ref{esym} it is seen that the symmetry energy
for these two models start to  differ at the saturation density and this may
explain the transition density behaviour. However the transition pressure increases 
faster for NL$\delta$, and for $B^*=5\times 10^4$ NL$\rho$ and NL$\delta$ 
have the same transition pressure.  
For DDRH models, DDRH$\delta$ predicts smaller densities for $B^*<3.5\times
10^5$. The change of behaviour for the larger fields seems to be related with
the change of curvature of the symmetry energy at $\sim 0.12$ fm$^{-3}$, above
which the symmetry energy for DDRH$\delta$ becomes much stiffer. It is interesting 
to notice that for these models only above  $B^*=5\times 10^4$ do the transition 
pressure and density increase with respect to the B=0 values;\\
c) even though the proton fractions and transition densities may be similar, all models
predict different pressures at the transition. NLW models have larger
pressures for  $B^*<3\times10^5$, NL$\delta$ being the one with the larger
values. For the DDRH models and below  $B^*<3\times10^5$, it is the model
including $\delta$ which has the smallest pressure, about half the value calculated
with NL$\delta$. 

For $B^*=5\times 10^5$, the transition density
is above the saturation density, $\rho\sim 0.16-0.17$ fm$^{-3}$ for
NLW models and to $\rho\sim0.19-0.23$ fm$^{-3}$ for DDRH models. 
For this high field both models with the $\delta$ meson
predict larger transition densities, pressures and proton fractions due to the
stiffness of the symmetry energy at those densities.
\begin{table}[bht]
\caption{Same as Table~\ref{table4}, but with AMM included.}  
\begin{ruledtabular}
\label{table5}
\begin{tabular}{ c c c c c}
$B^{*}$ & Model  & $\rho^{\hbox{cross}}_b (\hbox{fm}^{-3}) $ & $Y_p$ & $P_{m}(\hbox{MeV}\hbox{fm}^{-3}) $ \\
\hline
$5\times10^{4}$  & NL$\rho$          & 0.079 & 0.079 & 0.570  \\
                              & NL$\delta$        & 0.079 & 0.075 & 0.558  \\
                              & TW                    & 0.078 & 0.097 & 0.606 \\
                              & DDRH$\delta$   & 0.080 & 0.098 & 0.498 \\
$\phantom{..}$ \\
$10^{5}$              & NL$\rho$          & 0.093 & 0.152 & 1.274  \\
                              & NL$\delta$        & 0.094 & 0.149 & 1.329  \\
                              & TW                    & 0.091 & 0.178 & 1.281 \\
                              & DDRH$\delta$   & 0.088 & 0.183 & 1.108 \\
$\phantom{..}$\\
$2\times10^{5}$ & NL$\rho$           & 0.091 & 0.153 & 1.206  \\
                              & NL$\delta$        & 0.108 & 0.260 & 1.819 \\
                              & TW                    & 0.092 & 0.305 & 1.355  \\
                              & DDRH$\delta$   & 0.088 & 0.313 & 1.198 \\
$\phantom{..}$\\
$5\times10^{5}$ & NL$\rho$           & 0.134 & 0.434 & 1.693  \\
                             & NL$\delta$         & 0.143 & 0.442 & 2.245 \\
                             & TW                     & 0.149 & 0.488 & 2.772 \\
                             & DDRH$\delta$    & 0.170 & 0.504 & 4.356 \\
\end{tabular}
\end{ruledtabular}
\end{table}

In Table~\ref{table5} we show the same data given in Table~\ref{table4} but
including the AMM in the calculation. The conclusions are similar: for NLW models, the
transition density and the corresponding pressure increase when the $\delta$-meson is included, 
whereas for DDRH models the inclusion of the isovector-scalar meson decreases 
the transition density and the corresponding pressure for $B^{*}<5\times 10^{5}$. 
In summary, the $\delta$ meson  gives rise to a larger crust for NLW models and, 
except to very large fields, to a  smaller crust with DDRH models.

\begin{figure}[t]
\vspace{.75cm}
\centering
\includegraphics[width=1.0\linewidth,angle=0]{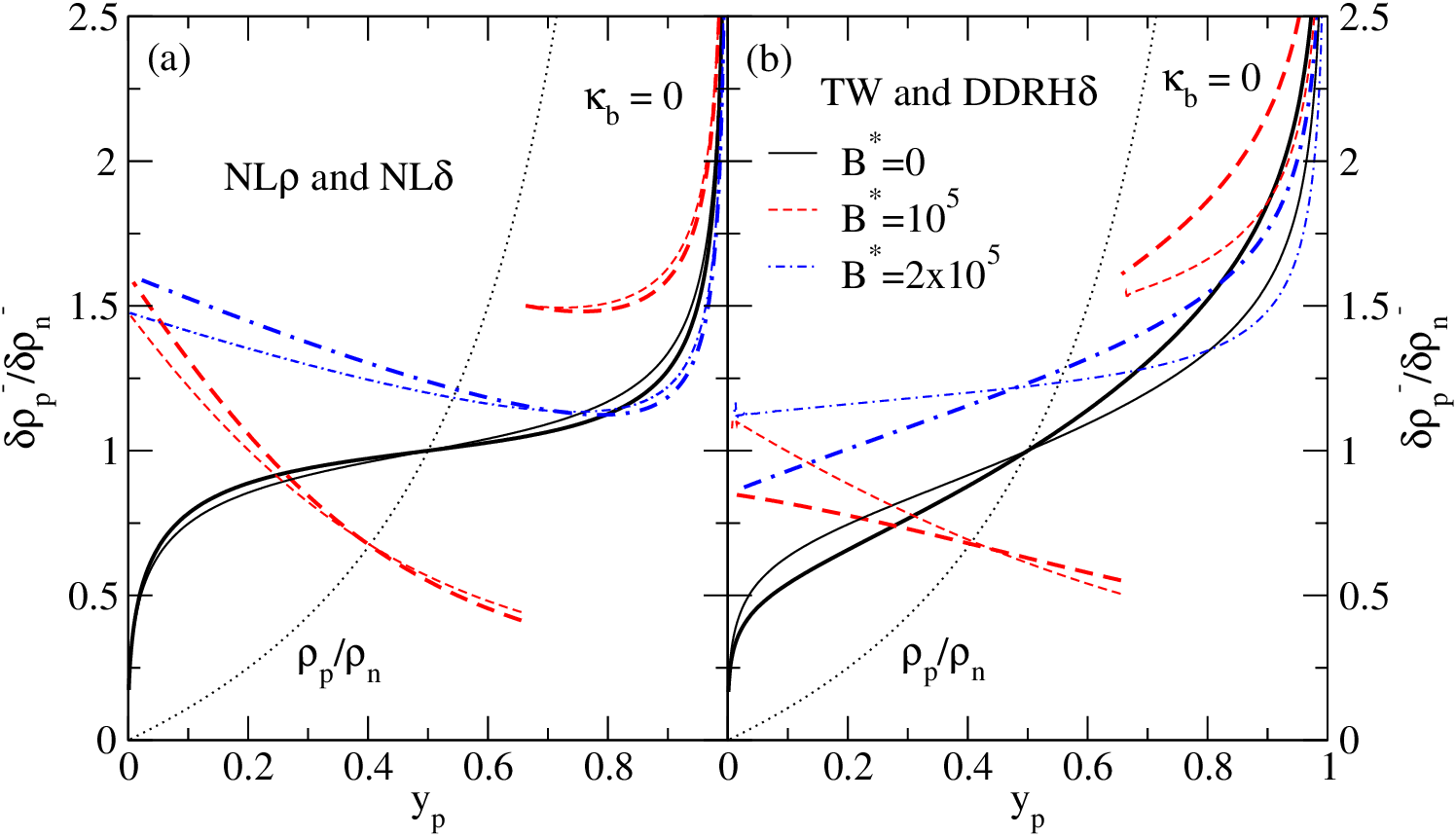}
\caption{(Color online) $\delta \rho^{-}_{p}/\delta \rho^{-}_{n}$ plotted as a
function of the proton fraction with  $\rho= 0.06
\hbox{fm}^{-3}$  for the NLW models (NL$\rho$ (thin lines) and NL$\delta$ (thick lines)) (left) 
and DDRH models (TW (thin lines) and DDRH$\delta$ (thick lines)) (right) and for several values of the magnetic fields without AMM. The fraction $\rho_p/\rho_n$ is given by the thin dotted line.}
\label{spddelta}
\end{figure}

The eigenvector associated with the eigenvalue of the free energy curvature
matrix  defines the direction of the instability and tells us how does the system
separate into a dense liquid and a gas phase. It was shown in \cite{abmp06,camille08} that
in the absence of the magnetic field the direction of instability favors the
reduction of the isospin asymmetry of the dense clusters of the system, and
increases the isospin asymmetry of the gas surrounding the clusters, the so
called distillation effect. This effect is represented in Fig.~\ref{spddelta}
where it is seen that for the $B=0$ curve the fraction $\delta
\rho^{-}_{p}/\delta \rho^{-}_{n}$ is larger than $\rho_p/\rho_n$ below $y_p=0.5$ 
and the other way round above. 
In this figure we show, respectively for NLW models (left) and DDRH models (right), 
the fraction $\delta \rho^{-}_{p}/\delta \rho^{-}_{n}$ as a function of $y_p$ 
for a fixed baryonic density, $\rho= 0.06 \hbox{ fm}^{-3}$, chosen inside the instability region.

For NLW models and for the two largest fields considered the spinodal 
region contains a single Landau level and the curve varies smoothly starting at  $\delta
\rho^{-}_{p}/\delta \rho^{-}_{n}\sim1.5$ for NL$\rho$ and $\sim 1.62$ for NL$\delta$. 
We point out the very large value of this fraction, always above 1. 
The magnetic field favors a strong increase of the proton fraction.
 For $y_p>0.5$, NL$\rho$ and NL$\delta$ behave in a
similar way, while below this value the main difference is the larger $\delta
\rho^{-}_{p}/\delta \rho^{-}_{n}$ for NL$\delta$ corresponding to a stronger distillation effect.

For $B^*=10^5$ the spinodal has two bands, see Fig.~\ref{spznlrdpn}, 
corresponding to the occupation of the
first two Landau levels. The transition from one to the other is clearly seen
with a large discontinuity of $\delta\rho^{-}_{p}/\delta \rho^{-}_{n}$ at $y_p\sim 0.7$. 
Above this $y_p$ value the
curve behaves like the previous ones. However for $y_p<0.7$ the behavior is
quite different: the curve decreases from the value at $y_p$=0, which is
independent of the magnitude of the  magnetic field, to a value much smaller
than the corresponding value of the fraction $\rho_p/\rho_n$. The same
behaviour occurs for NL$\rho$ and NL$\delta$. The fluctuations will not drive the system
out of the first Landau level and therefore the larger the proton fraction, the
closer the system comes to the top of the band and the smaller are the allowed
proton fluctuations. For $y_p>0.7$ or for the larger magnetic fields the
Landau levels are only partially filled and the fluctuations will never drive
the system out of the corresponding Landau level.
In summary,  the effect of the $\delta$-meson on the instability region of NLW models is to reduce 
the strength of the distillation effect of neutron-rich matter.

For DDRH models the inclusion of the isovector scalar meson reduces the 
strength of the distillation effect, which is not so dramatic in these models. 
Including the AMM similar conclusions are drawn. The AMM favors still larger 
proton fluctuations because neutron polarization stiffens the EOS.

In conclusion, we have studied the influence of the isovector scalar meson on the 
instabilities of stellar matter under very strong magnetic fields. The fields considered are 
much stronger than the strongest field measured at the surface of a magnetar which 
is $B^*\sim 10^2$ for SGR 1806-20 \cite{sgr}. However, the magnetic fields
in the interior of neutrons stars could be  larger and the present work shows how 
fields of the order of $B^*=10^{18}-10^{19}$ could affect the inner crust of a compact star. 
According to the scalar virial theorem~\cite{lai91} the maximum magnetic energy 
could be comparable to the gravitational energy in an equilibrium configuration, 
which would correspond to a value of the order of $\sim 10^{18}$ G.

We have considered  four relativistic nuclear models: two models with constant couplings (NL$\rho$
and NL$\delta$) and two models with density dependent couplings (TW and DDRH$\delta$). Two of the models include the $\delta$-meson. 
For all the models, we have determined the spinodal surface from the curvature
matrix of the free energy for different magnitudes of the magnetic field. 
It had already been shown~\cite{aziz09} that the instability region could be divided in to several bands
according to magnitude of the magnetic field and the number of  the Landau
levels occupied and that  the presence of the magnetic field would generally increase the
instability region.

We have seen that the inclusion of the $\delta$ meson increases (reduces) the size of the thermodynamic instability zone for very neutron-rich matter for the NLW (DDRH) models when
compared with the spinodal obtained without the $\delta$-meson. These results
reflect themselves on the extension of the crust of a compact star under a strong magnetic field.

By making a rough estimation of the transition density at the
inner crust of a compact star under a strong magnetic field from the crossing
of the EOS with the thermodynamic spinodal, we have shown that the transition
density and associated pressure increases, for NLW models, with the inclusion of the $\delta$
meson. On the other hand, for DDRH models the influence of the $\delta$-meson
is to decrease the transition density and associated pressure for fields
$B^{*}<5\times10^5$.  DDRH$\delta$ is predicting the smallest transition
densities, almost half the value obtained for NL$\delta$.

If we consider fields $\sim 10^{18}$ G ($B^*\sim 5\times 10^4$), as indicated by the scalar virial theorem, we may take the following conclusions: a) for conventional pulsars TW would predict a larger transition pressure and, therefore, larger fractional moment of inertia in the crust than all the other models, about twice as large; b) for $B^*\sim 5\times 10^4$, and taking into account AMM,  TW would predict just a small increase of the  fractional moment of inertia in the crust, $\sim 15$\%, the NLW models predict an increase of 120\% and 220\%, respectively, with and without $\delta$-meson, while DDRH$\delta$ would predict an increase of 80\%. If the AMM are not
considered, within  TW the fractional moment of inertia in the crust would be smaller than for conventional pulsars, while for DDRH$\delta$ there would be a slight increase of 20\% and for the NLW models an increase of $\sim$ 100\%. The different behaviours in magnetars and conventional pulsars predicted by
the different models might be a possibility to impose stronger constraints on
the EOS of nuclear matter.

We have also investigated the direction of instability. 
If the first Landau level is only partially occupied the density fluctuations
are such that the system evolves for a state with dense clusters very proton
rich immersed in a proton poor gas. A larger proton fraction is favored
energetically due to the degeneracy of the Landau levels. 
The $\delta$ meson will only reduce slightly the strength of this 
distillation effect for DDRH models and increase it for NLW models.
For particles occupying an almost complete Landau level, 
proton fluctuations are smaller or forbidden  and  an anti-distillation 
effect results with a decrease of the proton fraction of the dense clusters.

\begin{acknowledgments}
This work was partially supported by FEDER and FCT (Portugal) under Grant SFRH/BPD/14831/2003 
and Projects PTDC/FP/64707/2006 and CERN/FP/83505/2008. 

\end{acknowledgments}


\begin{thebibliography}{34}

\bibitem{usov}  V. V. Usov, Nature 357, 472 (1992).
\bibitem{pacz} B. Paczy\'nski, Acta Astron. 42, 145 (1992).	      	
\bibitem{duncan92}  Christopher Thompson and Robert C. Duncan,
Astrophys. J. L9, 392 (1992)
\bibitem{harding} A. K. Harding and D. Lai, Rep. Prog. Phys. 69, 2631 (2006).

\bibitem{sgr} 
SGR/APX online Catalogue,
http://www.physics.mcgill.ca/\verb ~ pulsar/magnetar/main.html

\bibitem{kouve} 
C. Kouveliotou, S. Dieter, T. Strohmayer, J. van Paradijs, G.J. Fishman, C.A. Meegan, K. Hurley, Nature 393, 235 (1998).

\bibitem{duncan}  Christopher Thompson and Robert C. Duncan,
Astrophys. J. 408, 194 (1993)
\bibitem{pethick94} D. G. Ravenhall and C. J. Pethick, Astrophys. J. 424, 846 (1994).

\bibitem{link99} Bennett Link, Richard I. Epstein, and James M. Lattimer, Phys. Rev. Lett. {\bf 83}, 3362 (1999).

\bibitem{kaspi} Victoria M. Kaspi,	Astrophys. Space Sci.  308,
1 (2007).

\bibitem{dib} Rim Dib, Victoria M. Kaspi, and Fotis Gavriil, Astrophys. J. 673, 1044 (2008).

\bibitem{aziz09} 
A. Rabhi, C  Provid\^encia, and J. da  Provid\^encia, Phys. Rev C 79, 015804 (2009).

\bibitem{camille08} Camille Ducoin, Constan\c ca Provid\^encia, Alexandre M. Santos, Lucilia
Brito, Philippe Chomaz, Phys. Rev. C 78, 055801 (2008).

\bibitem{kubis97}
S. Kubis and M. Kutschera, Phys. Lett. B399, 191 (1997)

\bibitem{liu} 
B. Liu, V. Greco, V. Baran, M. Colonna, and M. Di Toro, Phys. Rev. C 65, 045201 (2002).

\bibitem{mp04} 
D. P. Menezes and  C. Provid\^encia, Phys. Rev. C 70, 058801 (2004).

\bibitem{tw} 
S. Typel and H. H. Wolter, Nucl. Phys. A656, 331 (1999).

\bibitem{gaitanos} 
T. Gaitanos, M. Di Toro, S. Typel, V. Baran, C. Fuchs, V. Greco,  and H. H. Wolter, Nucl. Phys. A732, 24 (2004).

\bibitem{aziz08} 
A. Rabhi, C  Provid\^encia, and J. da  Provid\^encia, J. Phys. G: Nucl. Part. Phys. 35,125201 (2008).

\bibitem{fuchs} 
C. Fuchs, H. Lenske, and H. H. Wolter, Phys. Rev. C 52, 3043 (1995).

\bibitem{abmp04}  
S. S. Avancini,  L. Brito,  D. P. Menezes, and C. Provid\^encia,
Phys. Rev. C  70, 015203  (2004).

\bibitem{Bar03} 
H. M\"uller and B. D. Serot, Phys. Rev. C 52, 2072 (1995)
V. Baran, M. Colonna, M. Di Toro, and A. B. Larionov, Nucl. Phys. A632, 287 (1998).
\bibitem{marg03} J. Margueron and P. Chomaz, Phys. Rev. C {\bf 67}, 041602(R) 
(2003).

\bibitem{pasta2} S.S. Avancini, L. P. Brito,  J.R. Marinelli,  D.P. Menezes,
M.M.W. Moraes, C. Provid\^encia and A. M. Santos,  Phys. Rev. {\bf C 79}, 035804 (2009). 

\bibitem{pasta1} S.S. Avancini, D.P. Menezes, M.D. Alloy, J.R. Marinelli, 
M.M.W. Moraes and C. Provid\^encia,  Phys. Rev. {\bf C 78}, 015802 (2008). 
\bibitem{bao} J. Xu, L.W. Chen, B.A. Li and H.R. Ma, arXiv:0807.4477v1 
[nucl-th].

\bibitem{abmp06} 
S. S. Avancini,  L. Brito,  Ph. Chomaz,  D. P. Menezes, and  C. Provid\^encia, Phys. Rev. C 74, 024317 (2006).

\bibitem{lai91} Dong Lai and Stuart L. Shapiro, Astrophys. J. 383, 745 (1991)


\end{thebibliography}
\end{document}